\documentclass{elsart}

\usepackage{natbib}

\usepackage{epsfig}

\usepackage{amssymb}

\begin{document}

\begin{frontmatter}

\title{Status of non-Riemannian cosmology}

\author{Dirk Puetzfeld}
\address{Department of Physics and Astronomy, Iowa State University, Ames, IA 50011, USA}
\ead{dpuetz@iastate.edu}
\ead[url]{www.thp.uni-koeln.de/$\sim$dp}
\thanks[talk]{Extended version of a talk given at the 6th UCLA Symposium on ``Sources and Detection of Dark Matter and Dark Energy in the Universe'', February 18-20, 2004, Marina del Rey, CA, USA.}
\begin{abstract}
We provide a brief chronological guide to the literature on non-Riemannian cosmological models. Developments in this field are traced back to the early seventies and are given in table form.
\end{abstract}


\end{frontmatter}

\section{Introduction}
\label{sec_introduction}
Nowadays, cosmology is a prospering and rapidly changing field of physics. One could say that cosmology has entered a {\it golden age} with a wealth of new experimental data available, cf. the other contributions in this proceedings. In addition theoretical efforts have led to what we call the {\it standard cosmological model} or {\it cosmological concordance model}. Within this model we are able to explain most of the observations. But, maybe as a trade-off for simplicity, we have to introduce concepts like dark matter and dark energy within this picture. Unfortunately, so far there has been {\it no} direct detection of a dark matter particle and there seems to be {\it no} deeper theoretical justification for the large amount of dark energy. This is clearly an embarrassing situation because we cannot tell what over 90\% of the universe is really made of.
 
Since cosmology combines concepts of many different fields of physics there exist several distinct approaches to find a remedy for this unsatisfactory situation. One of them is to change the spacetime geometry and therewith the underlying gravity theory. In this report we want to focus on the history of non-Riemannian cosmological models, which we track back to the early seventies. For reviews covering non-Riemannian gravity theories the reader is referred to \cite{PhysRep}, \cite{Blagojevic}, \cite{Hammond1} and references therein.  

\section{Classification of non-Riemannian models}

\begin{figure}
\begin{center}
\setlength{\unitlength}{1mm}
\begin{picture}(140,120)
\epsfig{file=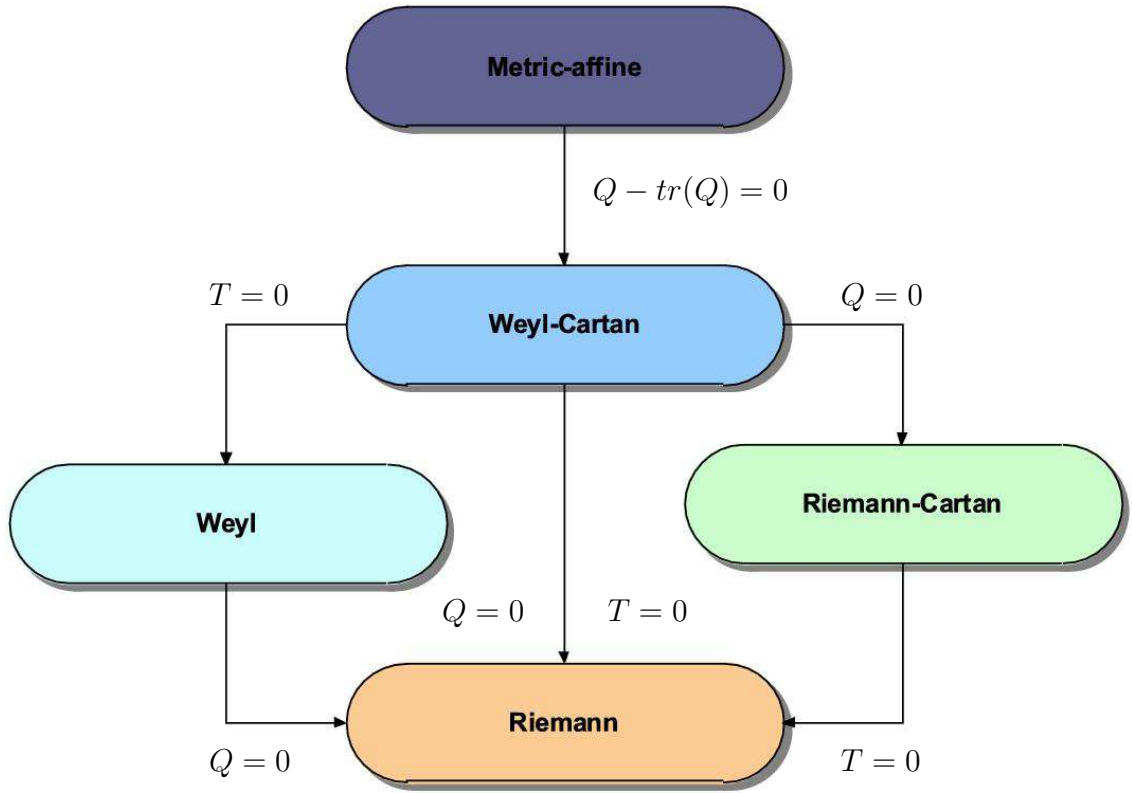,width=150mm}
\put(-72,79){{$Q-tr(Q)=0$}}
\put(-39,65){{$Q=0$}}
\put(-39,3){{$T=0$}}
\put(-123,65){{$T=0$}}
\put(-123,3){{$Q=0$}}
\put(-70,23){{$T=0$}}
\put(-92,23){{$Q=0$}}
\end{picture}\\
\setlength{\unitlength}{1pt}
\end{center}
\caption{Classification scheme for different spacetime types in terms of two of the field strengths of metric-affine gravity, i.e. nonmetricity $Q$ and torsion $T$.}
\label{figure_spacetime_types}
\end{figure}

One usually talks about non-Riemannian spacetimes as soon as the connection $\Gamma$, which plays a fundamental role in the parallel transport of geometrical objects, is no longer given by the metric compatible Christoffel connection that one encounters in Einstein's theory of general relativity (GR). The idea to consider more general connections and to think of the connection as an independent object, which is not necessarily tied to the metric, goes back to the works of \cite{Weyl} and \cite{Cartan}. Without going into detail we only mention that the advent of local gauge theories in the 1950s led to a renewed interest in the torsion $T$, i.e.\ the new antisymmetric piece of the connection introduced by Cartan, which may be related to spin. Nowadays nearly all non-Riemannian models may be classified within the language of metric-affine gravity (MAG), cf.\ \cite{Hehl1976} and \cite{PhysRep}. MAG represents a gauge theoretical formulation of a gravity theory and can be viewed as a generalization of Poincar\'{e} gauge theory (PGT). In contrast to Riemann-Cartan spacetime the connection in a metric-affine spacetime is no longer metric compatible, i.e.\ the covariant derivative of the metric does not vanish. The field related to this violation of the metricity condition is called nonmetricity $Q$. In figure \ref{figure_spacetime_types} we depict how one arrives at different spacetime types by switching off torsion $T$ and nonmetricity $Q$. A very general Lagrangian for MAG, encompassing more than twenty free parameters and thereby most of the models considered up to this date, has been suggested by \cite{Exact2}. 

In this review only the cosmological models which belong to the so-called nonsymmetric gravity theory (NGT), cf. \cite{Moffat1979}, do not fit into the framework of MAG since NGT also allows for antisymmetric metrics. Nonsymmetric metrics were already studied by \cite{Einstein} in their search for a unified theory for gravity and electromagnetism (see \cite{Goenner} for a comprehensive review of unified field theories). Although this unification was not successful the idea to use more general metrics in order to formulate a generalized theory of gravity persisted.

In the following we use the term non-Riemannian cosmology (NRC) synonymously for all cosmological models which are no longer tied to a Riemannian spacetime structure. 

\section{Cosmological models}\label{sec_cosmological_models}

In tables \ref{table_history1}--\ref{table_history4} we collected (in chronological order) the works\footnote{Since we collected all of the references in these four tables we refrain to cite individual works in this section.} on NRC beginning in the early 1970s. To our knowledge this is the most extensive compilation of works which is solely devoted to the subject of non-Riemannian {\it cosmological} models. We hope that this (very) condensed compendium proves to be useful for cosmologists who want to get a rough idea of what has been achieved in this field. Of course our compilation is not complete, especially several articles which were published in Russian journals had to be omitted since they were out of our reach. 

Most of the early cosmological models were based on Einstein-Cartan theory. Investigations mainly revolved around the construction of exact solutions and the question of whether or not an initial singularity can be avoided in such models. In the 1980s more general types of Lagrangians were considered. The inclusion of quadratic terms in the Lagrangian, leading to dynamical degrees of freedom, was mainly motivated by the framework of PGT and led to new classes of exact solutions.

The advent of the inflationary model led to a flood of works which tried to mimic or justify this new idea within different non-Riemannian scenarios. Till the end of the 1990s most of the works in NRC were focused on the description of the early stages of the universe. This bias can mainly be ascribed to the estimates for the new spin-spin contact interaction encountered in Einstein-Cartan theory. This interaction shows up at extremely high energy densities and might therefore play only a crucial role at Planck time scale. 

This is not to say that the impact of non-Riemannian effects is limited to the early universe, since one has to keep in mind that even tiny changes in the expansion history have profound consequences on the outcome of cosmological tests, such as primordial nucleosynthesis or the magnitude-redshift relation. In recent works some of these tests have been used to constrain the parameters in several non-Riemannian scenarios and thereby the presence of non-Riemannian quantities in the late stages of the universe. 

\section{Conclusion \& Outlook}

Challenging experimental results, like the (non) detection of dark energy and dark matter, are often the precursors of a shifting of scientific paradigms. Non-Riemannian gravity theories seem to offer an ideal playground for cosmologists and their search for an explanation of dark matter and dark energy.

Maybe cosmology might yield the definitive clue for the presence of non-Riemannian structures in our universe thereby utilizing a nearly 85 year old theoretical idea. 

\paragraph*{Acknowledgments}
The author is grateful to M.\ Daniel, J.\ Garecki, R.T.\ Hammond, F.W.\ Hehl, J.W.\ Moffat, Y.N.\ Obukhov, M.\ Pohl, P.\ Savaria, E.\ Scholz, I.L.\ Shapiro, and the organizers of this symposium, especially D.L.\ MacLaughlan-Dumes for her kind support. 
\paragraph*{Acronyms} Non-Riemannian cosmology (NRC), metric-affine gravity (MAG), nonsymmetric gravity theory (NGT), Poincar\'{e} gauge theory (PGT), general relativity (GR), Friedmann-Lema\^{\i}tre-Robertson-Walker (FLRW).   \\
\begin{table}
\caption{Brief history of non-Riemannian cosmological models}
\label{table_history1}
\begin{tabular}{llp{9cm}}
\hline
\hline
Reference & &Description\\
\hline
{\it \cite{Kopczynski1}}&&Spherically-symmetric solutions with torsion in Einstein-Cartan theory, singularity avoidance. \\
{\it \cite{Trautman}}&&Flat exact dust solution with spin in Einstein-Cartan theory, singularity avoidance, minimal radius well above Planck scale.  \\
{\it \cite{Kopczynski2}}&&Exact solutions for Bianchi I model in Einstein-Cartan theory, singularity avoidance only in special cases.\\
{\it \cite{Stewart}}&&Exact solution for Bianchi I model in Einstein-Cartan theory, singularity avoidance only in special cases, model compatible with the one of \cite{Kopczynski2}.\\
{\it \cite{Tafel}}&&Shows that non-singular Bianchi I-VIII models in Einstein-Cartan theory are possible, exact solutions for type I and V.\\
{\it \cite{Hehl1}}&&Generalized singularity theorem for Einstein-Cartan theory.\\
{\it \cite{Raychaudhuri}}&&Bianchi I model in Einstein-Cartan theory with magnetic field, singularity avoidance.\\
{\it \cite{Kerlick}}&&Initial singularity in Einstein-Cartan models rather enhanced than avoided if one takes Dirac field as source for metric and torsion.\\
{\it \cite{Kerlick1}}&&Investigates `bouncing' behavior of solutions in theories with torsion.\\
{\it \cite{Kunstatter}}&&Anisotropic vacuum solution within NGT.\\
{\it \cite{Tsamparlis0}}&&Consequences for torsion from certain symmetry (cosmological principle) restrictions in a Riemann-Cartan space.\\
{\it \cite{Minkevich1}}&&Generalized FLRW solution for a theory with torsion and quadratic Lagrangian, singularity avoidance.\\
{\it \cite{Tsamparlis}}&&Cosmological model within Einstein-Cartan theory, field equations of the Friedmann type with effective pressure and effective energy density.\\
{\it \cite{Minkevich2}}&&Early de Sitter type solutions in a theory with torsion.\\
{\it \cite{Nurgaliev}}&&Static cosmological solution in Einstein-Cartan theory.\\
{\it \cite{Canale}}&&Einstein-de Sitter solutions within PGT.\\
{\it \cite{Goenner1}}&&Class of exact homogeneous and isotropic solutions for 10 parameter PGT Lagrangian.\\
\hline
\hline
\end{tabular}
\end{table}

\begin{table}
\caption{Brief history of non-Riemannian cosmological models (continued)}
\label{table_history2}
\begin{tabular}{llp{9cm}}
\hline
\hline
{\it \cite{Smalley0}, \cite{Smalley1}}&&Cosmological model with spin-fluid and G\"{o}del metric in Einstein-Cartan theory.\\
{\it \cite{Buchbinder}}&&Nonsingular model with torsion, solution of field equations for effective action yields inflationary behavior.\\
{\it \cite{Garecki1}}&&Exact solutions for a model with torsion, Lagrangian with quadratic curvature and quadratic torsion terms.\\
{\it \cite{Demianski}}&&Inflationary solutions in Einstein-Cartan theory, singularity avoidance.\\
{\it \cite{Minkowski}}&&Exact solutions for an Einstein-Cartan model with spatially homogeneous torsion.\\
{\it \cite{Gasperini0}}&&Cosmological model within Einstein-Cartan theory, inflationary phase triggered by spin, singularity avoidance.\\
{\it \cite{Obukhov0}}&&Variational theory for the Weyssenhoff fluid in Einstein-Cartan theory, class of exact anisotropic homogeneous solutions, subclass of it avoids singularity, contains most of the early models as a subclass.\\
{\it \cite{Garecki2}}&&Qualitative discussion of the model proposed in \cite{Garecki1}, analysis of its singularities.\\
{\it \cite{Ritis}}&&Einstein-Cartan model with scalar field, singularity avoidance, inflationary solutions.\\
{\it \cite{Fennelly}}&&Inflationary solutions for Bianchi I models in Einstein-Cartan theory.\\
{\it \cite{Assad}}&&Inflationary solutions in Einstein-Cartan theory.\\
{\it \cite{Kao}}&&Inflationary solutions in a Weyl invariant scenario.\\
{\it \cite{Obukhov1}}&&Inflation triggered by a spin non-linearity in the matter Lagrangian.\\
{\it \cite{Tresguerres}}&&First cosmological model in Weyl-Cartan spacetime, exact solutions for torsion and nonmetricity.\\
{\it \cite{Chatterjee}}&&Exact inflationary solution for a model with additional quadratic torsion terms in the Einstein-Cartan Lagrangian.\\ 
{\it \cite{Poberii}}&&Inflationary solutions driven by nonmetricity.\\
{\it \cite{Garecki3}}&&Exact solutions for a restricted model within PGT, singularity avoidance, see also \cite{Garecki4}.\\
{\it \cite{Wolf}}&&Exacts solutions for model with effective pressure and effective energy density due to spin contributions, singularity avoidance.\\ 
\hline
\hline
\end{tabular}
\end{table}

\begin{table}
\caption{Brief history of non-Riemannian cosmological models (continued)}
\label{table_history3}
\begin{tabular}{llp{9cm}}
\hline
\hline
{\it \cite{Minkevich3}}&&Qualitative analysis of some exact cosmological solutions  in Weyl-Cartan spacetime.\\
{\it \cite{Obukhov}}&&First cosmological model within the triplet ansatz of MAG.\\
{\it \cite{Oliveira}}&&Exact solutions for a Weyl invariant scenario, inflationary solutions, singularity avoidance.\\
{\it \cite{Moffat1}}&&Homogeneous isotropic models within NGT reduce to FLRW models, NGT field equations for small antisymmetric field.\\
{\it \cite{Maroto}}&& String inspired model with dilaton and torsion, inflationary solutions due to torsion.\\
{\it \cite{Palle0}}&& Some speculations about formal analogies of cosmological models in Einstein-Cartan theory.\\
{\it \cite{Savaria}}&&Exact anisotropic vacuum and perfect fluid solutions within NGT, discussion of birefringence effects in NGT.\\
{\it \cite{TuckerWang}}&&Exact solutions within a triplet model, qualitative discussion of a new dark matter coupling related to the Proca charge of particles.\\
{\it \cite{Gasperini}}&&Avoidance of the initial singularity due to spin-torsion interactions in Einstein-Cartan theory, comparison to model in which the local Lorentz symmetry is broken.\\
{\it \cite{Minkevich4}}&&Field equations for the homogeneous and isotropic case within MAG, exact solutions for restricted Lagrangian in Weyl-Cartan spacetime.\\
{\it \cite{Capozziello0}}&&Scalar perturbations within an Einstein-Cartan model with a restricted type of torsion.\\
{\it \cite{Capozziello1}}&&Helicity flip probability induced by torsion, points out cosmological applications.\\
{\it \cite{Brueggen}}&&Estimates possible effects of torsion on primordial nucleosynthesis (based on work of \cite{Hammond}).\\
{\it \cite{Palle1}}&&Density fluctuations in Einstein-Cartan theory (based on the work of \cite{Obukhov0}).\\
{\it \cite{Andrade}}&&Some speculations about torsion effects in a dilaton model.\\
{\it \cite{PuetzTres}}&&Exact solutions and qualitative analysis of a model in Weyl-Cartan spacetime, cosmological constant term emerges naturally.\\
\hline
\hline
\end{tabular}
\end{table}

\begin{table}
\caption{Brief history of non-Riemannian cosmological models (continued)}
\label{table_history4}
\begin{tabular}{llp{9cm}}
\hline
\hline
{\it \cite{Moffat2}}&&NGT cosmological model, possible consequences for dark matter and dark energy sketched, no exact solution of the field equations.\\
{\it \cite{Puetzfeld1}}&&Exact solutions of a model in Weyl-Cartan spacetime, qualitative discussion of the solutions.\\
{\it \cite{Capozziello2}}&&Model with effective energy density and pressure due to torsion contributions, torsion plays the role of quintessence.\\
{\it \cite{Shapiro}}&&Model with totally antisymmetric torsion, exact solutions for the flat case, singularity avoidance.\\
{\it \cite{Puetzfeld2}}&&Parameter constraints for the Weyl-Cartan model from \cite{Puetzfeld1}, makes use of a combined SN Ia data set.\\
{\it \cite{Minkevich5}}&&Numerical analysis of the field equations from \cite{Minkevich4} with additional scalar field.\\
{\it \cite{Capozziello3}}&&Model with effective energy density and pressure due to torsion, torsion mimics cosmological constant, fits to SN Ia, SZE, and X-Ray data.\\
{\it \cite{Babourova}}&&Derivation of a model in Weyl-Cartan spacetime (partially resembles the model of \cite{Obukhov} and the model of \cite{Puetzfeld1}).\\
{\it \cite{Vereshchagin}}&&Qualitative analysis of the field equations from \cite{Minkevich4} with additional scalar field.\\
{\it \cite{Minkevich6}}&&Investigates `bouncing' behavior of the models from \cite{Minkevich5}, see also \cite{Minkevich7}.\\
{\it \cite{Puetzfeld4}}&&Comprehensive analysis of the SN Ia data within a Weyl-Cartan model, analysis valid for model of \cite{Obukhov} and model of \cite{Babourova}.\\
{\it \cite{Miritzis}}&&Weylian cosmological model, qualitative analysis of the field equations and asymptotic behavior.\\
{\it \cite{Moffat}}&&NGT cosmological model, possible consequences for dark matter and dark energy sketched, no exact solution of the field equations (enhanced version of \cite{Moffat2}).\\
{\it \cite{Scholz}}&&Weylian cosmological model, early estimates for some of the cosmological tests.\\
\hline
\hline
\end{tabular}
\end{table}

\begin{twocolumn}

\end{twocolumn}


\begin{thebibliography}{}

\bibitem[Assad \& Letelier(1990)]{Assad} Assad M.J.D., Letelier P.S., Phys. Lett. \textbf{A145} 74-78 (1990)

\bibitem[Babourova \& Frolov(2003)]{Babourova} Babourova O.V., Frolov B.N.,  Class. Quantum Grav. \textbf{20} 1423-1442 (2003)  \texttt{gr-qc/0209077}

%\bibitem[B\"auerle \& Haneveld(1983)]{Baeuerle} B\"auerle G.G.A., Haneveld C.J.,  Physica \textbf{A121} 541-554 (1983)  

\bibitem[Blagojevi\'{c}(2002)]{Blagojevic} Blagojevi\'{c} M.: \textit{Gravitation and gauge symmetries.} IOP Publishing, Bristol and Philadelphia (2002)

\bibitem[Br\"{u}ggen(1999)]{Brueggen} Br\"{u}ggen M., Gen. Rel. Grav. \textbf{31} 1935-1939 (1999)

\bibitem[Buchbinder et al.(1985)]{Buchbinder} Buchbinder I.L., Odintsov S.D., Shapiro I.L., Phys. Lett. \textbf{B162} 92-96 (1985)

\bibitem[Canale(1984)]{Canale} Canale A., de Ritis R., Tarantino C., Phys. Lett. \textbf{A100} 178-181 (1984)

\bibitem[Capozziello \& Stornaiolo(1999)]{Capozziello0} Capozziello S., Stornaiolo C., Nuov. Cim. \textbf{113B} 879-886 (1998)

\bibitem[Capozziello et al.(1999)]{Capozziello1} Capozziello S. et al., Europhys. Lett. \textbf{46} 710-715 (1998)

\bibitem[Capozziello(2002)]{Capozziello2} Capozziello S., Mod. Phys. Lett. \textbf{A25} 1621-1626 (2002)

\bibitem[Capozziello et al.(2003)]{Capozziello3} Capozziello S. et al., Int. J. Mod. \textbf{D12} 381-394 (2003)

\bibitem[Cartan(1922)]{Cartan} Cartan \'{E}., Math. Zeit. \textbf{2} 384 (1918)

\bibitem[Chatterjee \& Bhattacharya(1993)]{Chatterjee} Chatterjee P., Bhattacharya B., Mod. Phys. Lett. \textbf{A8} 2249-2257 (1993)

\bibitem[Demianski et al.(1986)]{Demianski} Demianski M. et al., Phys. Lett. \textbf{A116} 13-16 (1986)

\bibitem[Einstein \& Straus(1946)]{Einstein} Einstein A., Straus E.G., Ann. Math. \textbf{47} 731-741 (1946)

\bibitem[Fennelly et al.(1988)]{Fennelly} Fennelly A.J., Bradas J.C., Smalley L.L., Phys. Lett. \textbf{A129} 195-200 (1988)

\bibitem[Garcia de Andrade(1999)]{Andrade} Garcia de Andrade L.C., Phys. Lett. \textbf{B468} 28-30 (1999)

\bibitem[Garecki(1985)]{Garecki1} Garecki J., Acta Phys. Plon. \textbf{B16} 699-713 (1985)

\bibitem[Garecki(1987)]{Garecki2} Garecki J., Acta Phys. Plon. \textbf{B18} 147-161 (1987)

\bibitem[Garecki(1993)]{Garecki3} Garecki J., Gen. Rel. Grav. \textbf{25} 257-265 (1993)

\bibitem[Garecki(1995)]{Garecki4} Garecki J., Int. J. Theo. Phys. \textbf{34} 2307-2319 (1995)

\bibitem[Gasperini(1986)]{Gasperini0} Gasperini M., Phys. Rev. Lett. \textbf{56} 2873-2876 (1986)

\bibitem[Gasperini(1998)]{Gasperini} Gasperini M., Gen. Rel. Grav. \textbf{30} 1703-1709 (1998)

\bibitem[Goenner \& M\"{u}ller-Hoissen(1984)]{Goenner1} Goenner H., M\"{u}ller-Hoissen F., Class. Quantum Grav. \textbf{1} 651-672 (1984)

\bibitem[Goenner(2004)]{Goenner} Goenner H.F.M., http://relativity. livingreviews.org/lrr-2004-2 (2004) 

\bibitem[Hammond(1996)]{Hammond} Hammond R.T., Class. Quantum Grav. \textbf{13} 1691-1697 (1996)

\bibitem[Hammond(2002)]{Hammond1} Hammond R.T., Rep. Prog. Phys. \textbf{65} 599-649 (2002)

\bibitem[Hehl et al.(1974)]{Hehl1} Hehl F.W., von der Heyde P., Kerlick G.D., Phys. Rev. \textbf{D10} 1066-1069 (1974)

\bibitem[Hehl et al.(1976)]{Hehl1976} Hehl F.W., von der Heyde P., Kerlick G.D., Nester J.M., Rev. Mod. Phys. \textbf{48} 393-416 (1976)

\bibitem[Hehl et al.(1995)]{PhysRep}   Hehl F.W., McCrea J.D., Mielke E.W., Ne\'{}eman Y., Phys. Rep. \textbf{258} 1-171 (1995)

\bibitem[Hehl \& Mac\'{\i}as(1999)]{Exact2}  Hehl F.W., Mac\'{\i}as A., Int. J. Mod. Phys. \textbf{D8} 399-416 (1999) 

\bibitem[Kao(1990)]{Kao} Kao W.F., Phys. Lett. \textbf{A149} 76-78 (1990)

\bibitem[Kerlick(1975)]{Kerlick} Kerlick G.D., Phys. Rev. \textbf{D12} 3004-3006 (1975)

\bibitem[Kerlick(1976)]{Kerlick1} Kerlick G.D., Annals Phys. \textbf{99} 127-141 (1976) 

\bibitem[Kopczy\'{n}ski(1972)]{Kopczynski1} Kopczy\'{n}ski W., Phys. Lett. \textbf{A39} 219-220 (1972)

\bibitem[Kopczy\'{n}ski(1973)]{Kopczynski2} Kopczy\'{n}ski W., Phys. Lett. \textbf{A43} 63-64 (1973)

\bibitem[Kunstatter et al.(1979)]{Kunstatter} Kunstatter G., Moffat J.W., Savaria P., Phys. Rev. \textbf{D19} 3559-3561 (1979)

\bibitem[Maroto \& Shapiro(1997)]{Maroto} Maroto A.L., Shapiro I.L., Phys. Lett. \textbf{B414} 34-44 (1997)

\bibitem[Minkevich(1980)]{Minkevich1} Minkevich A.V., Phys. Lett. \textbf{A80} 232-234 (1980) 

\bibitem[Minkevich(1983)]{Minkevich2} Minkevich A.V., Phys. Lett. \textbf{A95} 422-424 (1983)

\bibitem[Minkevich \& Nemenman(1995)]{Minkevich3} Minkevich  A.V.,  Nemenman I.M., Class. Quantum Grav. \textbf{12} 1259-1265 (1995) 

\bibitem[Minkevich \& Garkun(1999)]{Minkevich4} Minkevich A.V., Garkun A.S., Grav. Cosm. \textbf{5} 115 \texttt{gr-qc/9805007} 

\bibitem[Minkevich(2002)]{Minkevich5} Minkevich A.V., Int. J. Mod. Phys. \textbf{A17} 4441-4449 (2002) 

\bibitem[Minkevich(2003a)]{Minkevich6} Minkevich A.V., Nonlin. Phenom. Complex. Syst. \textbf{6} 824-832 (2003) \texttt{gr-qc/0307026}

\bibitem[Minkevich(2003b)]{Minkevich7} Minkevich A.V., \texttt{gr-qc/0312068}

\bibitem[Minkowski(1986)]{Minkowski} Minkowski P., Phys. Lett. \textbf{B173} 247-250 (1986)

\bibitem[Miritzis(2004)]{Miritzis} Miritzis J., \texttt{gr-qc/0402039}  

\bibitem[Moffat(1979)]{Moffat1979} Moffat J.W., Phys. Rev. \textbf{19} 3554 (1979)

\bibitem[Moffat(1997)]{Moffat1} Moffat J.W., \texttt{astro-ph/9704300}

\bibitem[Moffat(2001)]{Moffat2} Moffat J.W., \texttt{astro-ph/0108201}

\bibitem[Moffat(2004)]{Moffat} Moffat J.W., \texttt{astro-ph/0403266}

\bibitem[Nurgaliev \& Ponomariev(1983)]{Nurgaliev} Nurgaliev I.S., Ponomariev W.N., Phys. Lett. \textbf{B130} 378-379 (1983)

\bibitem[Obukhov \& Korotky(1987)]{Obukhov0} Obukhov Y.N., Korotky V.A., Class. Quantum Grav. \textbf{4} 1633-1657 (1987)

\bibitem[Obukhov(1993)]{Obukhov1} Obukhov Y.N., Phys. Lett. \textbf{A182} 214-216 (1993)

\bibitem[Obukhov et al.(1997)]{Obukhov}  Obukhov Y.N., Vlachynsky E.J., Esser W., Hehl F.W., Phys. Rev. \textbf{D56} 7769-7778 (1997) 

\bibitem[de Oliveira et al.(1997)]{Oliveira} de Oliveira H.P., Salim J.M., Sautu S.L., Class. Quantum Grav.\textbf{14} 2833-2843 (1997) 

\bibitem[Palle(1997)]{Palle0} Palle D., Nuov. Cim. \textbf{111B} 671 (1997)

\bibitem[Palle(1999)]{Palle1} Palle D., Nuov. Cim. \textbf{114B} 853-860 (1999)

\bibitem[Poberii(1994)]{Poberii} Poberii E.A., Helv. Phys. Acta \textbf{67} 745-752 (1994)

\bibitem[Puetzfeld \& Tresguerres(2001)]{PuetzTres}  Puetzfeld D., Tresguerres R., Class. Quantum Grav. \textbf{18} 667-693 (2001) 

\bibitem[Puetzfeld(2002a)]{Puetzfeld1}  Puetzfeld D., Class. Quantum Grav. \textbf{19} 3363-3280 (2002)

\bibitem[Puetzfeld(2002b)]{Puetzfeld2}  Puetzfeld D., Class. Quantum Grav. \textbf{19} 4463-4482 (2002)

\bibitem[Puetzfeld \& Chen(2004)]{Puetzfeld4}  Puetzfeld D., Chen X., Class. Quantum Grav. \textbf{21} 2703-2722 (2004) \texttt{gr-qc/0402026}  

\bibitem[Raychaudhuri(1975)]{Raychaudhuri} Raychaudhuri A.K., Phys. Rev. \textbf{D12} 952-955 (1975)

\bibitem[de Ritis et al.(1988)]{Ritis} de Ritis R., Scudellaro P., Stornaiolo C., Phys. Lett. \textbf{A126} 389-392 (1988)

\bibitem[Savaria(1997)]{Savaria} Savaria P., \texttt{gr-qc/9711004} 

\bibitem[Scholz(2004)]{Scholz} Scholz E., \texttt{astro-ph/0403446}

\bibitem[Shapiro(2002)]{Shapiro} Shapiro I.L., Phys. Rep. \textbf{357} 113-213 (2002)

\bibitem[Smalley(1985a)]{Smalley0} Smalley L.L., Phys. Rev. \textbf{D12} 3124-3127 (1985)

\bibitem[Smalley(1985b)]{Smalley1} Smalley L.L., Phys. Lett. \textbf{A113} 463-466 (1985)

\bibitem[Stewart \& H\'{a}j\'{\i}\v{c}ek(1973)]{Stewart}Stewart J., H\'{a}j\'{\i}\v{c}ek P., Nature (Phys. Sci.) \textbf{244} 96 (1973)

\bibitem[Tafel(1973)]{Tafel}  Tafel J., Phys. Lett. \textbf{A45} 341-342 (1973)

\bibitem[Trautman(1973)]{Trautman} Trautman A., Nature (Phys. Sci.) \textbf{242} 7-8 (1973)

\bibitem[Tresguerres(1993)]{Tresguerres} Tresguerres R., Proc. Relativity in general, Salas, Asturias, (Spain), Sept. 7-10 (1993) 407-413

\bibitem[Tsamparlis(1979)]{Tsamparlis0} Tsamparlis M., Phys. Lett. \textbf{A75} 27-28 (1979) 

\bibitem[Tsamparlis(1981)]{Tsamparlis} Tsamparlis M., Phys. Rev. \textbf{D24} 1451-1457 (1981)

\bibitem[Tucker \& Wang(1998)]{TuckerWang} Tucker R.W., Wang C., Class. Quantum Grav. \textbf{15} 933-954 (1998) 

\bibitem[Vereshchagin(2003)]{Vereshchagin} Vereshchagin G.V., Int. J. Mod. Phys. \textbf{D12} 1487-1498 (2003)

\bibitem[Weyl(1918)]{Weyl} Weyl H., Math. Zeit. \textbf{2} 384 (1918)

\bibitem[Wolf(1995)]{Wolf} Wolf C., Gen. Rel. Grav. \textbf{27} 1031-1042 (1995)
\end{thebibliography}
\end{document}